# Disorder-free sputtering method on graphene

Xue Peng Qiu[1], Young Jun Shin[1], Jing Niu[1], Narayanapillai Kulothungasagaran[1], Gopinadhan Kalon[1,2], Caiyu Qiu[3], Ting Yu[2,3], and Hyunsoo Yang[1,2,a]

[1]Department of Electrical and Computer Engineering, National University of Singapore, 117576, Singapore

[2]Graphene Research Centre, National University of Singapore, 117546, Singapore

[3]Division of Physics and Applied Physics, School of Physical and Mathematical Sciences, Nanyang Technological University, 637616, Singapore

Deposition of various materials onto graphene without causing any disorder is highly desirable for graphene applications. Especially, sputtering is a versatile technique to deposit various metals and insulators for spintronics, and indium tin oxide to make transparent devices. However, the sputtering process causes damage to graphene because of high energy sputtered atoms. By flipping the substrate and using a high Ar pressure, we demonstrate that the level of damage to graphene can be reduced or eliminated in dc, rf, and reactive sputtering processes.

[a] E-mail address: eleyang@nus.edu.sg

Graphene has attracted enormous scientific and engineering interest due to its superior electronic, optical, thermal, mechanical properties[1-9]. The important properties of graphene include a high intrinsic charge carrier mobility and a tunability of its properties by applying a gate bias voltage. In order to make top-gate graphene field effect transistors (G-FET), a high-k dielectric gate material needs to be deposited onto graphene without damage, because the mobility of graphene is outstanding when graphene is pristine.[10,11] The other interesting application is spintronics, in which an extremely long spin diffusion length or spin lifetime is preferred possibly due to the weak spin-orbit coupling strength.[12,13] In order to achieve efficient spin injection, high quality ferromagnetic contacts and tunnel barriers are essential without degrading the intrinsic properties of graphene during the deposition. All of these emphasize the great importance of a non-destructive deposition technique in constructing graphene devices for both fundamental studies and practical applications.

Although graphene is the strongest material ever known[4,5], but since graphene is only one monolayer of carbon atoms packed into a two-dimensional honeycomb lattice, the C-C bonds are easy to be broken during the deposition. Therefore, only thermal evaporation is known to be a non-destructive method of deposition onto graphene, which greatly limits the choice of materials in graphene devices. Sputtering is a widely used technique in industries for hard disk drives and for indium tin oxide coating as well as in the spintronics community. However, high energy atom bombardment induces a large amount of disorder into graphene, which greatly degrades the graphene's property. There have been a few studies based on sputtering onto graphene and all of them exhibited significant disorder[14-16].

In this letter, we propose a defect-free sputter deposition technique onto graphene by flipping the sample at high Ar pressure during the deposition in order to reduce the energy of incident sputtering atoms. We investigate different sputtering techniques such as dc, rf, and reactive sputtering, and study the level of damage of graphene by Raman spectroscopy. By flipping the samples and sputtering at high Ar pressure (20 mTorr), we successfully reduce the degree of disorder to a negligible level.

The graphene samples were prepared by micromechanical exfoliation of Kish graphite and subsequently transferred to a highly p-doped Si substrate, which has a layer of 300 nm thick $SiO_2$. Single layer graphene (SLG) was identified by an optical microscope and confirmed using a 532 nm Raman spectrometer[17-19]. We have first tried various methods to deposit thin films onto SLG and studied the level of damage of graphene by Raman spectroscopy. 3 nm $SiO_2$, 2 nm $TiO_2$, 2 nm Cr, 2 nm Cr and 2 nm Cu has been deposited onto SLG by plasma enhanced chemical vapor deposition (PECVD), pulsed laser deposition (PLD), thermal evaporation, e-beam evaporation, and sputtering with their typical deposition parameters, respectively. For sputtering, 3 mTorr Ar pressure and 60 W were used. As seen in Fig. 1, only thermal evaporation gives rise to a negligible D peak after the deposition. A significant D peak appears with the other methods, such as PLD, e-beam evaporation, PECVD, and sputtering. Among these methods, PLD and sputtering induce most significant disorder. The Raman spectra of graphene mainly consist of three peaks: D peak (~ 1350 $cm^{-1}$), G peak (~ 1580 $cm^{-1}$), and 2D peak (~ 2680 $cm^{-1}$). The G peak is due to the bond stretching of the pairs of $sp^2$ carbon atom. The D peak is due to the ring-breathing modes and is induced by disorder. The 2D line corresponds to a high-energy second-order process and is observed even in the absence of the D peak. The shape of the 2D peak and the ratio of intensity between the G peak ($I_G$) and 2D peak ($I_{2D}$) are used to discriminate SLG from multilayered graphene. From the shape and position of these three peaks and the ratio $I_D/I_G$, Ferrari and Robertson introduced a three stage model of disorder in carbon materials, which allows to simply assess the Raman spectra of graphene: the early stage leads to nanocrystalline graphite (nc-G phase) from crystalline graphite, the second stage is low tetragonal amorphous carbon (a-C phase), and the third stage is high $sp^3$ tetrahedral amorphous carbon (ta-C phase)[20]. In the following, we refer to these three stages to quantify the impact of deposition on the structural quality of graphene sheets. According to this model, for e-beam evaporation and PECVD, the disorder level is moderate and the amorphization is at the first stage. A second stage amorphization has been occurred to graphene with the PLD and sputtering processes.

We have further deposited the materials onto graphene using sputtering in two configurations; one is the normal the other is the flipping configuration. The schematic of two configurations is shown in Fig. 2. In the normal configuration, the SLG faces the sputter targets. In the flipping case, since the samples are flipped as the backside of the SLG samples faces the targets, the energy of atom bombardment can be greatly reduced especially at a high Ar pressure, as shown in Fig. 2(b), when the materials are deposited onto the flipped sample surface. 4 nm $Co_{70}Fe_{30}$ and 2 nm Al is deposited onto SLG by dc sputtering at a power of 60 W. 3 nm MgO is deposited by rf sputtering at 120 W, and 1 nm MgO is deposited by dc reactive sputtering at 60 W in a mixture of 30 sccm Ar and 1 sccm $O_2$ gas. All the deposition pressures are set to 20 mTorr which is much higher than a typical value of 3 mTorr in the sputtering process. The distance from the target to substrate is fixed to 30 cm. The purpose of using a high Ar pressure of 20 mTorr is to increase the atoms collision probability. Therefore, more atoms can be condensate onto the flipped sample surface, and the energy of atoms reduces when they reach the graphene surface. The deposition rate reduces in the flipping configuration. For example, the deposition rate of CoFe, Al, and MgO (rf) is 6.3 nm, 5.6 nm, and 6.3 nm per hour, respectively, in the normal configuration, while it is 3.5 nm, 2.5 nm, and 0.5 nm per hour in the flipping configuration, respectively. The deposition rate of reactive sputtered MgO using the flipping method is 1 nm per hour.

Figure 3(a) shows the Raman spectra after the deposition of 4 nm CoFe on SLG by two methods. With the normal deposition method, the appearance of the D peak indicates that the deposition of CoFe breaks the symmetry of graphene and induces disorder. However, the in-plane correlation length ($L_a$) is calculated to be 6.07 and the disorder level is still good for certain graphene applications due to the deposition in high Ar pressure.[21] This shows that high Ar pressure (20 mTorr) for sputtering can greatly reduce the disorder level as compared to the result of low Ar pressure (3 mTorr) sputtering as shown in Fig 1. The level of damage of graphene can be further reduced once the flipping method is utilized together with high Ar pressure. The Raman spectrum using the flipping method shows a negligible D peak, while the G and 2D peaks preserve their shapes, indicating the suitability of proposed dc sputtering method onto graphene.

The Raman spectra after the deposition of 2 nm Al on SLG are shown in Fig. 3(b). The result is similar to the case of CoFe. No disorder is seen from the spectra by using the flipping method, however, slight disorder appears in the normal method. A common adopted method to form $AlO_x$ is to deposit Al and then subsequently oxidized it in atmosphere, pure $O_2$, or oxygen plasma.[18] Since $AlO_x$ is often used as a tunnel barrier for spintronic devices or a dielectric layer to apply gate bias, our work thus sheds light on future graphene applications via sputtering.

Simply reducing the sputtering power is not helpful to reduce the damage level of graphene. For example, we reduce the CoFe deposition power from 60 to 23 W for the normal sputtering configuration, so that the deposition rate is the same to the flipping configuration. The D peak appears in the Raman spectra which is comparable with the one seen at 60 W in the same configuration (not shown). The improvement in disorder using the flipping method should mainly attribute to the reduced energy of atoms when they reach the sample surface rather than a slow deposition rate.

We have also deposited 3 nm of MgO on SLG by rf sputtering. As seen from Fig. 3(c), in the normal configuration, the shapes of the G and 2D peaks vanish which show a clear amorphization of graphene. With the flipping method, the D and G peaks can be still observed after the deposition, while the 2D peak disappears from the spectrum. Compared to the normal deposition configuration, the flipping method thus shows an improvement, but is not suitable for a high quality graphene device. This is due to the limitation of rf sputtering. In rf sputtering a high frequency ac voltage is applied between the ground of the sample holder and the target to discharge the target surface. The plasma of rf sputtering is more extended and $Ar^+$ ions are also present around the sample, therefore, rf sputtering has an enhanced ion bombardment which will induce large disorder onto graphene in both configurations. This can explain larger disorder of graphene due to rf deposited MgO as compared to $AlO_x$ obtained by dc sputtering from an Al target as reported previously.[14]

Reactive sputtering is an alternative method to deposit high quality tunnel barriers in a reactive gas mixture with Ar.[22] We have deposited 1 nm MgO onto graphene by dc reactive sputtering with the flipping method and the Raman spectra is shown in Fig. 3(d). A small D peak is observed, which comes from the oxygen plasma due to the oxygen gas mixture. By utilizing the proposed flipping method in high Ar pressure, a better quality of graphene is obtained after the oxide deposition as compared to the previous reports[14,16,23], which will enable to use various oxide materials in graphene devices by sputtering. Especially, highly spin filtering MgO tunnel barriers are of great importance for spintronic applications. By replacing the $O_2$ gas with the $N_2$ or other reactive gases, various nitrides and other materials can be explored with graphene using reactive sputtering.

We check the uniformity of deposited materials onto graphene by the proposed sputtering method. Figure 4 shows the atomic force microscopy (AFM) images of 4 nm CoFe and 2 nm Al on graphene which were deposited by the flipping method with 20 mTorr Ar pressure. The mean roughness of CoFe on graphene is 0.432 nm, while that of Al is 0.284 nm. The films show a good uniformity promising for practical applications.

In conclusion, we report a method for sputtering onto graphene without disorder by flipping the sample holder in a high Ar pressure. Negligible disorder is observed from the Raman spectra of graphene after the deposition of CoFe and Al by dc sputtering, and the AFM images show a small roughness of the films on graphene. MgO deposited by rf sputtering induces large disorder onto graphene due to its enhanced bombardment effect. However, reactive sputtering of MgO onto graphene leads to a small disorder compared to rf sputtering. The availability of sputtering for the deposition of various materials in the graphene system will pave a way of rigorous material engineering for various graphene applications.

This work is partially supported by the Singapore National Research Foundation under CRP Award No. NRF-CRP 4-2008-06 and "Novel 2D materials with tailored properties: beyond graphene" (R-144-000-295-281).

Figure captions

Figure 1. Raman spectra of SLG with the various deposition methods.

Figure 2. Schematic of sputtering deposition in the normal configuration with low Ar pressure (a) and the flipping configuration with high Ar pressure (b). The arrows show the trajectory of the sputtered atoms.

Figure 3. Raman spectra of SLG after dc sputtering of 4 nm CoFe (a) and 2 nm Al (b), rf sputtering of 3 nm MgO (c), and reactive sputtering of 1 nm MgO (d) with the normal (blue) and flipping (red) methods.

Figure 4. AFM images of CoFe (a, c) and Al (b, d) on graphene. (a) and (b) show the surface morphology over $1.5 \times 1.5$ µm$^2$. (c) and (d) show a line profile.

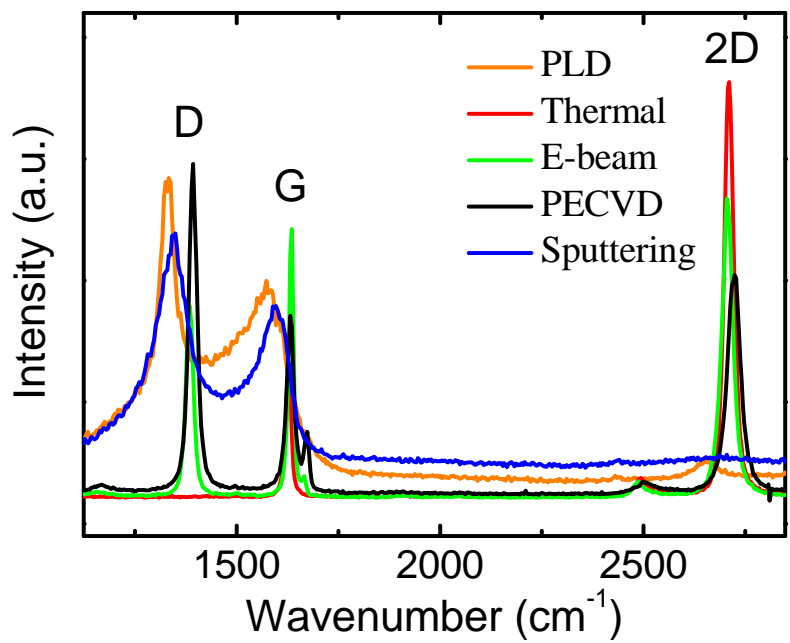

Figure 1.

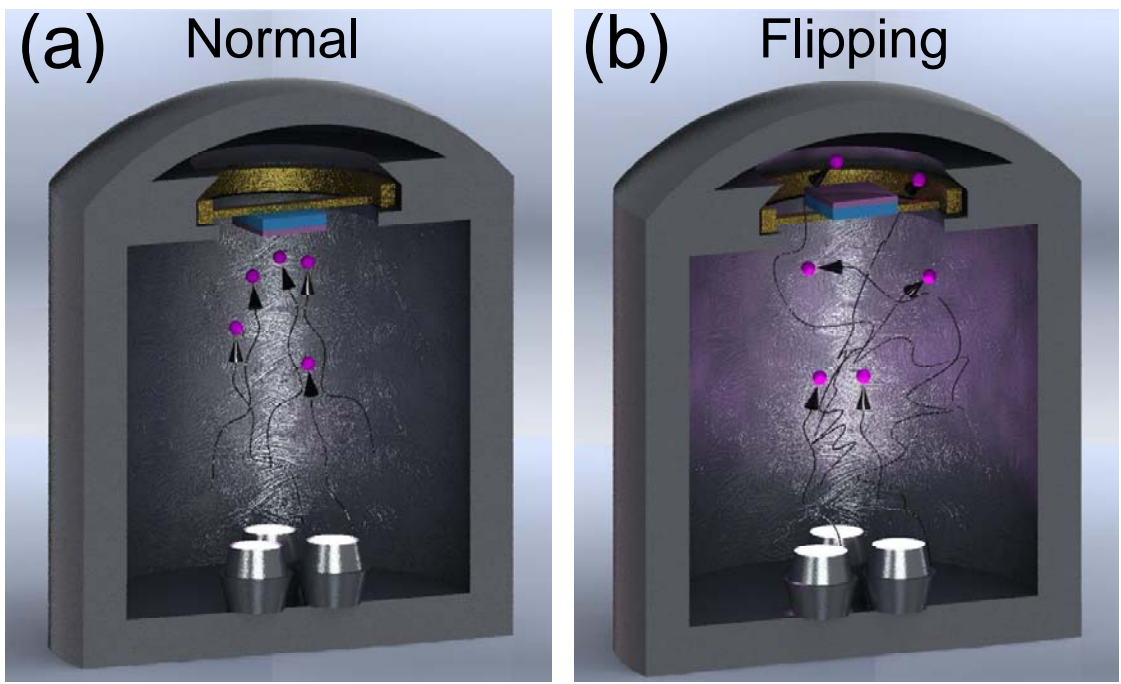

Figure 2.

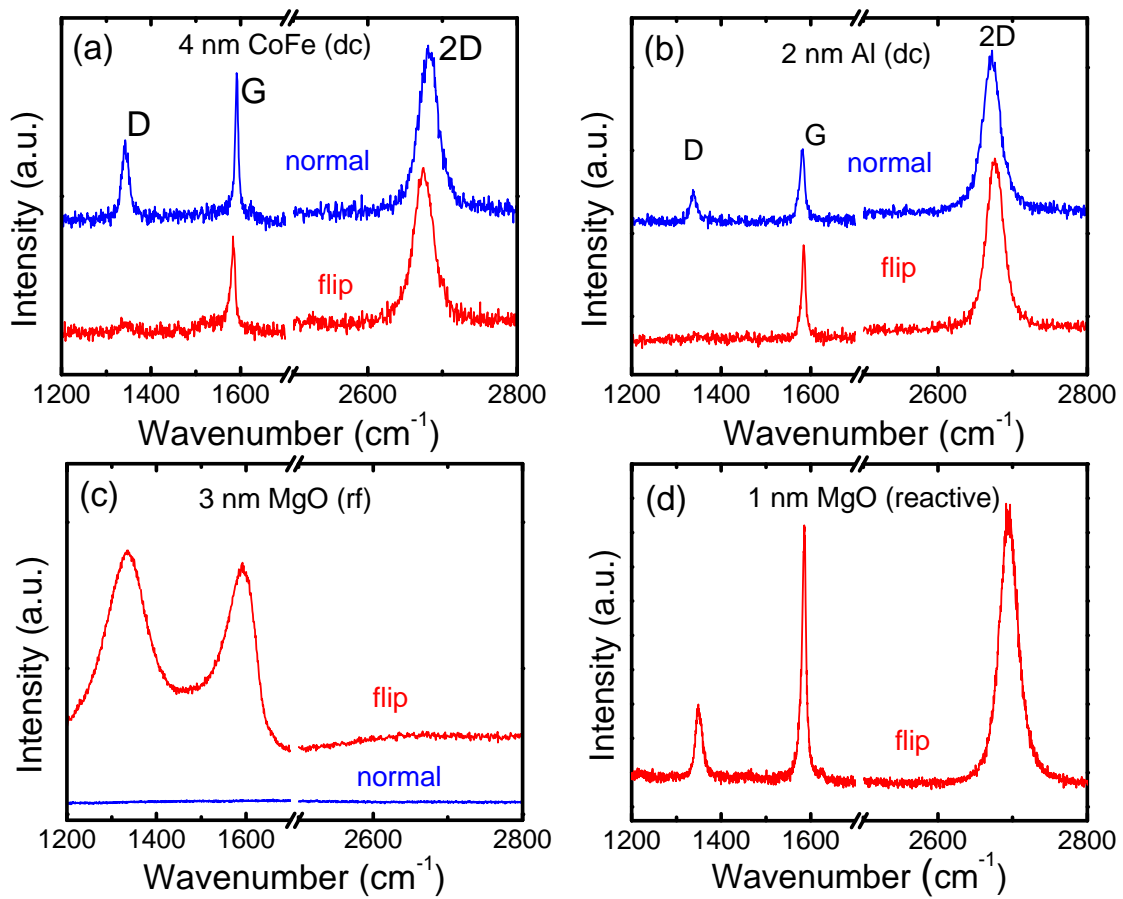

Figure 3.

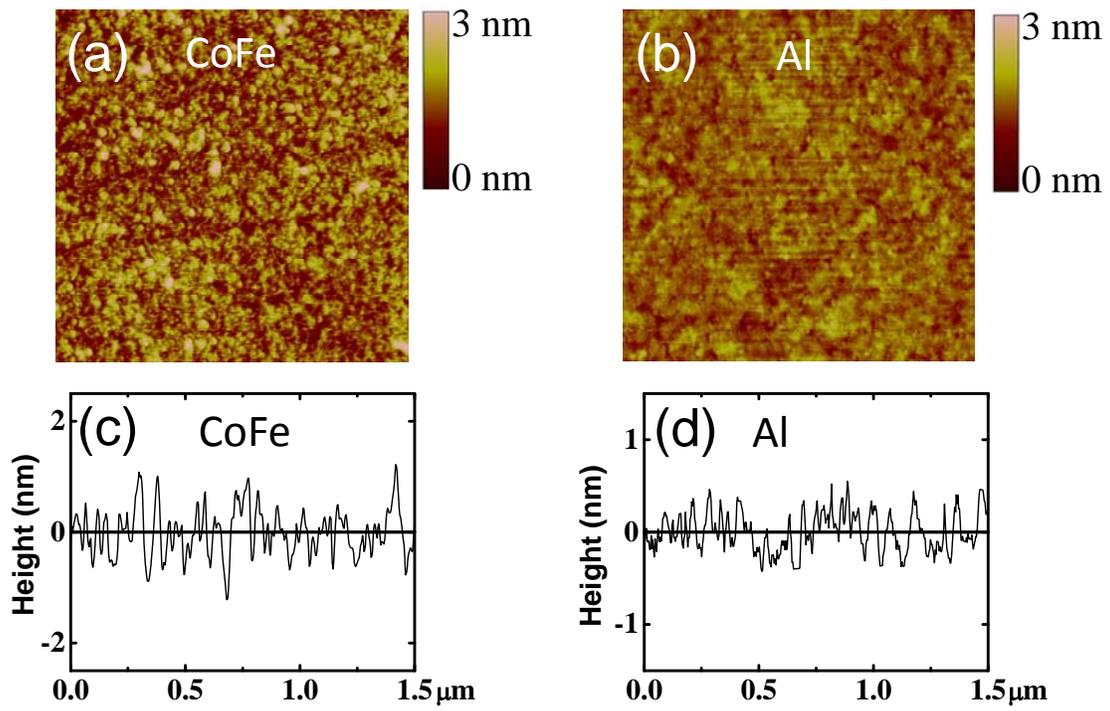

Figure 4.